\documentclass[smallextended]{svjour3}
\usepackage{mathptmx}
\usepackage{color}
\usepackage{amsmath}
\usepackage{amssymb}
\usepackage{graphics}
\usepackage[unicode=true,pdfusetitle,
 bookmarks=true,bookmarksnumbered=false,bookmarksopen=false,
 breaklinks=false,pdfborder={0 0 1},colorlinks=true]
 {hyperref}
 
\journalname{Gen. Rel. Grav.}

\begin{document}

\title{Effect of cosmic backreaction on the future evolution of an accelerating
universe}

\author{Nilok Bose \and 
        A. S. Majumdar
}

\institute{N. Bose \at 
S. N. Bose National Centre for Basic Sciences,
Block JD, Sector III, Salt Lake,
Kolkata 700098, India\\
\email{nilok@bose.res.in}
\and
A. S. Majumdar \at
S. N. Bose National Centre for Basic Sciences,
Block JD, Sector III, Salt Lake,
Kolkata 700098, India\\
\email{archan@bose.res.in}
}

\date{Received: date / Accepted: date}

\maketitle
 
\begin{abstract}
We investigate the effect of backreaction due to inhomogeneities on
the evolution of the present universe by considering a two-scale model
within the Buchert framework. Taking the observed present acceleration
of the universe as an essential input, we study the effect of inhomogeneities
in the future evolution. We find that the backreaction from inhomogeneities
causes the acceleration to slow down in the future for a range of
initial configurations and model parameters. The present acceleration
ensures formation of the cosmic event horizon, and our analysis brings
out how the effect of the event horizon could further curtail the
global acceleration, and even lead in certain cases to the emergence
of a future decelerating epoch. 

\keywords{Dark Energy, Cosmic Backreaction, Large Scale Structure}
\PACS{98.80.-k, 95.36.+x, 98.80.Es}
\end{abstract}

\section{Introduction}

The present acceleration of the Universe is well established observationally \cite{perlmutter},
but far from understood theoretically. The simplest possible explanation
provided by a cosmological constant is endowed with several conceptual
problems \cite{weinberg}. Alternative scenarios based on mechanisms
such as modification of the gravitational theory and invoking extra
fields invariably involve making unnatural assumptions on the model
parameters, though there is no dearth of innovative ideas to account
for the present acceleration \cite{sahni}. The present state of knowledge
offers no clear indication on the nature of the big rip that our Universe
seems to be headed for.

Observations of the large scale universe confirm the existence of
matter inhomogeneities up to the scales of superclusters of galaxies,
a feature that calls for at least an, in principle, modification of
the cosmological framework based on the assumption of a globally smooth
Friedmann\textendash{}Robertson\textendash{}Walker (FRW) metric. Taking
the global average of the Einstein tensor is unlikely to lead to the
same results as taking the average over all the different local metrics
and then computing the global Einstein tensor for a nonlinear theory
such as general relativity. This realization has lead to investigation
of the question of how backreaction originating from density inhomogeneities
could modify the evolution of the universe as described by the background
FRW metric at large scales.

In recent times there is an upsurge of interest on studying the effects
of inhomogeneities on the expansion of the Universe. The main obstacle
to this investigation is the difficulty of solving the Einstein equations
for an inhomogeneous matter distribution and calculating its effect
on the evolution of the Universe through tensorial averaging techniques.
Approaches have been developed to calculate the effects of inhomogeneous
matter distribution on the evolution of the Universe, such as Zalaletdinov's
fully covariant macroscopic gravity \cite{zala}; Buchert's approach
of averaging the scalar parts of Einstein's equations \cite{buchert1,buchert2};
 perturbation techniques proposed by Kolb et. al. \cite{kolb1} and the
light cone formalism for gauge invariant averaging \cite{gasperini}.
Based on the framework developed by Buchert it has been argued by
R\"{a}s\"{a}nen \cite{rasanen1} that backreaction from inhomogeneities from
the era of structure formation could lead to an accelerated expansion
of the Universe. The Buchert framework from a different perspective
developed by Wiltshire \cite{wiltshire1} also leads to an apparent
acceleration due to the different lapse of time in underdense and
overdense regions. Various other observable consequences of inhomogeneities
have been explored recently, such as their effect on dark energy
precision measurements \cite{ito} and the equation of state in dark energy 
models \cite{marra}. The scale of spherical
inhomogeneity around an observer in an accelerating universe 
may play an important role \cite{clarkson},
leading to the possible background dependence of the inferred nature of 
acceleration \cite{fleury}.

Backreaction from inhomogeneities provides an interesting platform
for investigating the issue of acceleration in the current epoch without
invoking additional physics, since the effects of backreaction gain
strength as the inhomogeneities develop into structures around the
present era. It needs to be mentioned here that the impact of inhomogeneities
on observables \cite{ishibashi,paranajpe} of an overall homogeneous
FRW model has been debated in the literature \cite{kolb2}. Similar
questions have also arisen with regard to the magnitude of backreaction
modulated by the effect of shear between overdense and underdense
regions \cite{mattson}. While further investigations with the input
of data from future observational probes are required for a conclusive
picture to emerge, recent studies \cite{buchert1,buchert2,buchert3,rasanen1,kolb1,wiltshire1,kolb2,weigand}
have provided a strong motivation for exploring further the role of
inhomogeneities in the evolution of the present Universe.

The backreaction scenario within the Buchert framework \cite{buchert1,buchert3,weigand}
has been recently studied by us from new perspective, \textit{viz.},
the effect of the event horizon on cosmological backreaction \cite{bose}.
The acceleration of the universe leads to a future event horizon from
beyond which it is not possible for any signal to reach us. The currently
accelerating epoch dictates the existence of an event horizon since
the transition from the previously matter-dominated decelerating expansion.
Since backreaction is evaluated from the global distribution of matter
inhomogeneities, the event horizon demarcates the spatial regions
which are causally connected to us, and hence impact the evolution
of our part of the Universe. By considering the Buchert framework
with the explicit presence of an event horizon during the present
accelerating era, our results \cite{bose} indicated the possibility
of a transition to a future decelerated era.

In the present paper we perform a comprehensive analysis of various
aspects of the above model for the future evolution of our universe
in the presence of cosmic inhomogeneities. We consider the universe
as a global domain $\mathcal{D}$ which is large enough to have a
scale of homogeneity to be associated with it. This global domain
$\mathcal{D}$ is then further partitioned into overdense and underdense
regions called $\mathcal{M}$ and $\mathcal{E}$ respectively, following
the approach in \cite{weigand}. In this setup we explore the
future evolution of the universe and then compare the results to those
that are obtained when the effect of the event horizon is taken into
account. In this manner we are able to demarcate the regions in parameter
space for which the effect of backreaction from inhomogeneities causes
the global acceleration to first slow down, and then ensures the onset
of another future decelerating era.

The paper is organized as follows. In Section 2 we briefly recapitulate
the essential details of the Buchert framework \cite{buchert1,buchert2,buchert3}
including the two-scale model of inhomogeneities presented in \cite{weigand}.
Next, in Section 3 we follow the approach of \cite{weigand} and investigate
the future evolution of the universe assuming its present stage of
global acceleration. We then illustrate the effects of the event horizon
by analyzing numerically the evolution equations in Section 4. Subsequently,
in Section 5 we perform a quantitative comparison of various dynamical
features of our model in the presence of an explicit event horizon
\cite{bose} with the future evolution as obtained within the standard
Buchert framework \cite{weigand}. Finally, we summarize our results
and make some concluding remarks in Section 6.

\section{The Backreaction Framework}

\subsection{Averaged Einstein equations}

In the framework developed by Buchert \cite{buchert1,buchert2,buchert3,weigand}
for the Universe filled with an irrotational fluid of dust, the space\textendash{}time
is foliated into flow-orthogonal hypersurfaces featuring the line-element
\begin{equation}
ds^{2}=-dt^{2}+g_{ij}dX^{i}dX^{j}
\end{equation}
where the proper time $t$ labels the hypersurfaces and $X^{i}$ are
Gaussian normal coordinates (locating free-falling fluid elements
or generalized fundamental observers) in the hypersurfaces, and $g^{ij}$
is the full inhomogeneous three metric of the hypersurfaces of constant
proper time. The framework is applicable to perfect fluid matter models.

For a compact spatial domain $\mathcal{D}$, comoving with the fluid,
there is one fundamental quantity characterizing it and that is its
volume. This volume is given by 
\begin{equation}
|\mathcal{D}|_{g}=\int_{\mathcal{D}}d\mu_{g}
\end{equation}
where $d\mu_{g}=\sqrt{^{(3)}g(t,X^{1},X^{2},X^{3})}dX^{1}dX^{2}dX^{3}$.
From the domain's volume one may define a scale-factor 
\begin{equation}
a_{\mathcal{D}}(t)=\left(\frac{|\mathcal{D}|_{g}}{|\mathcal{D}_{i}|_{g}}\right)^{1/3}
\end{equation}
that encodes the average stretch of all directions of the domain.

Using the Einstein equations, with a pressure-less fluid source, we
get the following equations \cite{buchert1,buchert3,weigand} 
\begin{eqnarray}
3\frac{\ddot{a}_{\mathcal{D}}}{a_{\mathcal{D}}} & = & -4\pi G\left\langle \rho\right\rangle _{\mathcal{D}}+\mathcal{Q}_{\mathcal{D}}+\Lambda\label{eq:1a}\\
3H_{\mathcal{D}}^{2} & = & 8\pi G\left\langle \rho\right\rangle _{\mathcal{D}}-\frac{1}{2}\mathcal{\left\langle R\right\rangle }_{\mathcal{D}}-\frac{1}{2}\mathcal{Q}_{\mathcal{D}}+\Lambda\label{eq:1b}\\
0 & = & \partial_{t}\left\langle \rho\right\rangle _{\mathcal{D}}+3H_{\mathcal{D}}\left\langle \rho\right\rangle _{\mathcal{D}}\label{eq:1c}
\end{eqnarray}
Here the average of the scalar quantities on the domain $\mathcal{D}$
is defined as, 
\begin{equation}
\left\langle f\right\rangle {}_{\mathcal{D}}(t)=\frac{\int_{\mathcal{D}}f(t,X^{1},X^{2},X^{3})d\mu_{g}}{\int_{\mathcal{D}}d\mu_{g}}=|\mathcal{D}|_{g}^{-1}\int_{\mathcal{D}}fd\mu_{g}\label{eq:2}
\end{equation}
and where $\rho$, $\mathcal{R}$ and $H_{\mathcal{D}}$ denote the
local matter density, the Ricci-scalar of the three-metric $g_{ij}$,
and the domain dependent Hubble rate $H_{\mathcal{D}}=\dot{a}_{\mathcal{D}}/a_{\mathcal{D}}$
respectively. The kinematical backreaction $\mathcal{Q_{D}}$ is defined
as 
\begin{equation}
\mathcal{Q_{D}}=\frac{2}{3}\left(\left\langle \theta^{2}\right\rangle _{\mathcal{D}}-\left\langle \theta\right\rangle _{\mathcal{D}}^{2}\right)-2\sigma_{\mathcal{D}}^{2}\label{eq:3}
\end{equation}
where $\theta$ is the local expansion rate and $\sigma^{2}=1/2\sigma_{ij}\sigma^{ij}$
is the squared rate of shear. It should be noted that $H_{\mathcal{D}}$
is defined as $H_{\mathcal{D}}=1/3\left\langle \theta\right\rangle _{\mathcal{D}}$.
$\mathcal{Q_{D}}$ encodes the departure from homogeneity and for
a homogeneous domain its value is zero.

One also has an integrability condition that is necessary to yield
\eqref{eq:1b} from \eqref{eq:1a} and that relation reads as 
\begin{equation}
\frac{1}{a_{\mathcal{D}}^{6}}\partial_{t}\left(a_{\mathcal{D}}^{6}\mathcal{Q}_{\mathcal{D}}\right)+\frac{1}{a_{\mathcal{D}}^{2}}\partial_{t}\left(a_{\mathcal{D}}^{2}\mathcal{\left\langle R\right\rangle }_{\mathcal{D}}\right)=0
\end{equation}
From this equation we see that the evolution of the backreaction term,
and hence extrinsic curvature inhomogeneities, is coupled to the average
intrinsic curvature. Unlike the FRW evolution equations where the
curvature term is restricted to an $a_{\mathcal{D}}^{-2}$ behaviour,
here it is more dynamical because it can be any function of $a_{\mathcal{D}}$.

\subsection{Separation formulae for arbitrary partitions}

The ``global'' domain $\mathcal{D}$ is assumed to be separated
into subregions $\mathcal{F}_{\ell}$ , which themselves consist of
elementary space entities $\mathcal{F}_{\ell}^{(\alpha)}$ that may
be associated with some averaging length scale. In mathematical terms
$\mathcal{D}=\cup_{\ell}\mathcal{F}_{\ell}$, where $\mathcal{F}_{\ell}=\cup_{\alpha}\mathcal{F}_{\ell}^{(\alpha)}$
and $\mathcal{F}_{\ell}^{(\alpha)}\cap\mathcal{F}_{m}^{(\beta)}=\emptyset$
for all $\alpha\neq\beta$ and $\ell\neq m$. The average of the scalar
valued function $f$ on the domain $\mathcal{D}$ \eqref{eq:2} may
then be split into the averages of $f$ on the subregions $\mathcal{F}_{\ell}$
in the form, 
\begin{equation}
\left\langle f\right\rangle _{\mathcal{D}}=\underset{\ell}{\sum}|\mathcal{D}|_{g}^{-1}\underset{\alpha}{\sum}\int_{\mathcal{F}_{\ell}^{(\alpha)}}fd\mu_{g}=\underset{\ell}{\sum}\lambda_{\ell}\left\langle f\right\rangle _{\mathcal{F}_{\ell}}
\end{equation}
where $\lambda_{\ell}=|\mathcal{F}_{\ell}|_{g}/|\mathcal{D}|_{g}$,
is the volume fraction of the subregion $\mathcal{F}_{\ell}$. The
above equation directly provides the expression for the separation
of the scalar quantities $\rho$, $\mathcal{R}$ and $H_{\mathcal{D}}=1/3\left\langle \theta\right\rangle _{\mathcal{D}}$.
However, $\mathcal{Q_{D}}$, as defined in \eqref{eq:3}, does not
split in such a simple manner due to the $\left\langle \theta\right\rangle _{\mathcal{D}}^{2}$
term. Instead the correct formula turns out to be 
\begin{equation}
\mathcal{Q_{D}}=\underset{\mathcal{D}}{\mathcal{\sum}}\lambda_{\ell}\mathcal{Q}_{\ell}+3\underset{\ell\neq m}{\sum}\lambda_{\ell}\lambda_{m}\left(H_{\ell}-H_{m}\right)^{2}\label{eq:4}
\end{equation}
where $\mathcal{Q}_{\ell}$ and $H_{\ell}$ are defined in $\mathcal{F}_{\ell}$
in the same way as $\mathcal{Q}_{\mathcal{D}}$ and $H_{\mathcal{D}}$
are defined in $\mathcal{D}$. The shear part $\left\langle \sigma^{2}\right\rangle _{\mathcal{F}_{\ell}}$
is completely absorbed in $\mathcal{Q}_{\ell}$ , whereas the variance
of the local expansion rates$\left\langle \theta^{2}\right\rangle _{\mathcal{D}}-\left\langle \theta\right\rangle _{\mathcal{D}}^{2}$
is partly contained in $\mathcal{Q}_{\ell}$ but also generates the
extra term $3\sum_{\ell\neq m}\lambda_{\ell}\lambda_{m}\left(H_{\ell}-H_{m}\right)^{2}$.
This is because the part of the variance that is present in $\mathcal{Q}_{\ell}$,
namely $\left\langle \theta^{2}\right\rangle _{\mathcal{\mathcal{F}_{\ell}}}-\left\langle \theta\right\rangle _{\mathcal{F}_{\ell}}^{2}$
only takes into account points inside $\mathcal{F}_{\ell}$. To restore
the variance that comes from combining points of $\mathcal{F}_{\ell}$
with others in $\mathcal{F}_{m}$, the extra term containing the averaged
Hubble rate emerges. Note here that the above formulation of the backreaction
holds in the case when there is no interaction between the overdense
and the underdense subregions.

Analogous to the scale-factor for the global domain, a scale-factor
$a_{\ell}$ for each of the subregions $\mathcal{F}_{\ell}$ can be
defined such that $|\mathcal{D}|_{g}=\sum_{\ell}|\mathcal{F}_{\ell}|_{g}$,
and hence, 
\begin{equation}
a_{\mathcal{D}}^{3}=\sum_{\ell}\lambda_{\ell_{i}}a_{\ell}^{3}\label{globscale}
\end{equation}
where $\lambda_{\ell_{i}}=|\mathcal{F}_{\ell_{i}}|_{g}/|\mathcal{D}_{i}|_{g}$
is the initial volume fraction of the subregion $\mathcal{F}_{\ell}$.
If we now twice differentiate this equation with respect to the foliation
time and use the result for $\dot{a}_{\ell}$ from \eqref{eq:1b},
we then get the expression that relates the acceleration of the global
domain to that of the sub-domains: 
\begin{equation}
\frac{\ddot{a}_{\mathcal{D}}}{a_{\mathcal{D}}}=\underset{\ell}{\sum}\lambda_{\ell}\frac{\ddot{a}_{\ell}(t)}{a_{\ell}(t)}+\underset{\ell\neq m}{\sum}\lambda_{\ell}\lambda_{m}\left(H_{\ell}-H_{m}\right)^{2}\label{eq:5}
\end{equation}
Following the simplifying assumption of \cite{weigand} (which
captures the essential physics), we work with only two subregions.
Clubbing those parts of $\mathcal{D}$ which consist of initial overdensity
as $\mathcal{M}$ (called `wall'), and those with initial underdensity
as $\mathcal{E}$ (called `void'), such that $\mathcal{D}=\mathcal{M}\cup\mathcal{E}$,
one obtains $H_{\mathcal{D}}=\lambda_{\mathcal{M}}H_{\mathcal{M}}+\lambda_{\mathcal{E}}H_{\mathcal{E}}$,
with similar expressions for $\left\langle \rho\right\rangle _{\mathcal{D}}$
and $\left\langle \mathcal{R}\right\rangle _{\mathcal{D}}$, and 
\begin{equation}
\frac{\ddot{a}_{\mathcal{D}}}{a_{\mathcal{D}}}=\lambda_{\mathcal{M}}\frac{\ddot{a}_{\mathcal{M}}}{a_{\mathcal{M}}}+\lambda_{\mathcal{E}}\frac{\ddot{a}_{\mathcal{E}}}{a_{\mathcal{E}}}+2\lambda_{\mathcal{M}}\lambda_{\mathcal{E}}(H_{\mathcal{M}}-H_{\mathcal{E}})^{2}\label{eq:6}
\end{equation}
Here $\sum_{\ell}\lambda_{\ell}=\lambda_{\mathcal{M}}+\lambda_{\mathcal{E}}=1$,
with $\lambda_{\mathcal{M}}=|\mathcal{M}|/|\mathcal{D}|$ and $\lambda_{\mathcal{E}}=|\mathcal{E}|/|\mathcal{D}|$.

\section{Future evolution within the Buchert framework}

We now try to see what happens to the evolution of the universe once
the present stage of acceleration sets in. Note, henceforth, we do
not need to necessarily assume that the acceleration is due to backreaction
\cite{rasanen1,weigand}. For the purpose of our present analysis,
it suffices to consider the observed accelerated phase of the universe
\cite{seikel} that could occur due to any of a variety of mechanisms
\cite{sahni}.

Since the global domain $\mathcal{D}$ is large enough for a scale
of homogeneity to be associated with it, one can write, 
\begin{equation}
|\mathcal{D}|_{g}=\int_{\mathcal{D}}\sqrt{-g}\, d^{3}X=f(r)a_{F}^{3}(t)
\end{equation}
where $f(r)$ is a function of the FRW comoving radial coordinate
$r$. It then follows that 
\begin{equation}
a_{\mathcal{D}}\approx\left(\frac{f(r)}{|\mathcal{D}_{i}|_{g}}\right)^{1/3}a_{F}\equiv c_{F}a_{F},
\end{equation}
and hence the volume average scale-factor $a_{\mathcal{D}}$ and the
FRW scale-factor $a_{F}$ are related by $a_{\mathcal{D}}\approx c_{F}a_{F}$,
where $c_{F}$ is a constant in time. Thus, $H_{F}\approx H_{\mathcal{D}}$,
where $H_{F}$ is the FRW Hubble parameter associated with $\mathcal{D}$.
Though in general $H_{\mathcal{D}}$ and $H_{F}$ could differ on
even large scales \cite{weigand}, the above approximation is valid
for small metric perturbations.

\begin{figure}
\centerline{\includegraphics{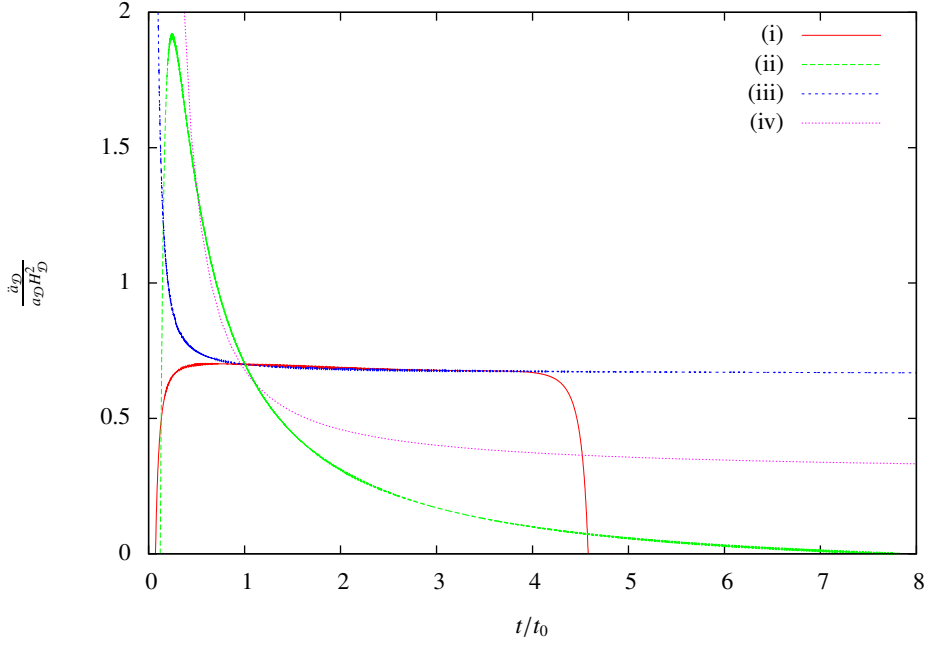} }

\caption{The dimensionless global acceleration parameter $\frac{\ddot{a}_{\mathcal{D}}}{a_{\mathcal{D}}H_{\mathcal{D}}^{2}}$,
plotted vs. time (in units of $t/t_{0}$ with $t_{0}$ being the current age of the universe
). The parameter values used are: (i) $\alpha=0.995$, $\beta=0.5$,
(ii) $\alpha=0.995$, $\beta=0.5,$ (iii) $\alpha=1.02$, $\beta=0.66$,
(iv) $\alpha=1.02$, $\beta=0.66$. (The curves (i) and (iii) correspond
to the case when an event horizon is included in the analysis in Section
4).}

\label{fig1} 
\end{figure}

Following the Buchert framework \cite{buchert1,weigand} as discussed
above, the global domain $\mathcal{D}$ is divided into a collection
of overdense regions $\mathcal{M}=\cup_{j}\mathcal{M}^{j}$, with
total volume $|\mathcal{M}|_{g}=\sum_{j}|\mathcal{M}^{j}|_{g}$ ,
and underdense regions $\mathcal{E}=\cup_{j}\mathcal{E}^{j}$ with
total volume $|\mathcal{E}|_{g}=\sum_{j}|\mathcal{E}^{j}|_{g}$. Assuming
that the scale-factors of the regions $\mathcal{E}^{j}$ and $\mathcal{M}^{j}$
are, respectively, given by $a_{\mathcal{E}_{j}}=c_{\mathcal{E}_{j}}t^{\alpha}$
and $a_{\mathcal{M}_{j}}=c_{\mathcal{M}_{j}}t^{\beta}$ where $\alpha$,
$\beta$, $c_{\mathcal{E}_{j}}$ and $c_{\mathcal{M}_{j}}$ are constants,
one has 
\begin{equation}
a_{\mathcal{E}}^{3}=c_{\mathcal{E}}^{3}t^{3\alpha};\>\>\>a_{\mathcal{M}}^{3}=c_{\mathcal{M}}^{3}t^{3\beta}\label{eq:8}
\end{equation}
where $c_{\mathcal{E}}^{3}=\frac{\sum_{j}c_{\mathcal{E}_{j}}^{3}|\mathcal{E}_{i}^{j}|_{g}}{|\mathcal{E}_{i}|_{g}}$
is a new constant, and similarly for $c_{\mathcal{M}}$. The volume
fraction of the subdomain $\mathcal{M}$ is given by $\lambda_{\mathcal{M}}=\frac{|\mathcal{M}|_{g}}{|\mathcal{D}|_{g}}$,
which can be rewritten in terms of the corresponding scale factors
as $\lambda_{\mathcal{M}}=\frac{a_{\mathcal{M}}^{3}|\mathcal{M}_{i}|_{g}}{a_{\mathcal{D}}^{3}|\mathcal{D}_{i}|_{g}}$.
We therefore find that the global acceleration equation \eqref{eq:6}
becomes 
\begin{eqnarray}
\frac{\ddot{a}_{\mathcal{D}}}{a_{\mathcal{D}}} & = & \frac{g_{\mathcal{M}_{h}}^{3}t^{3\beta}}{a_{\mathcal{D}}^{3}}\frac{\beta(\beta-1)}{t^{2}}+\left(1-\frac{g_{\mathcal{M}_{h}}^{3}t^{3\beta}}{a_{\mathcal{D}}^{3}}\right)\frac{\alpha(\alpha-1)}{t^{2}}\nonumber \\
 &  & +2\frac{g_{\mathcal{M}_{h}}^{3}t^{3\beta}}{a_{\mathcal{D}}^{3}}\left(1-\frac{g_{\mathcal{M}_{h}}^{3}t^{3\beta}}{a_{\mathcal{D}}^{3}}\right)\left(\frac{\beta}{t}-\frac{\alpha}{t}\right)^{2}\label{eq:19}
\end{eqnarray}
where $g_{\mathcal{M}_{h}}^{3}=\frac{\lambda_{\mathcal{M}_{0}}a_{\mathcal{D}_{0}}^{3}}{t_{0}^{3\beta}}$
is a constant. We obtain numerical solutions of the above equation
for various parameter values (see curves (ii) and (iv) of Fig.1).
The expansion factor $\beta$ for the overdense subdomain (wall) is
chosen to lie between $1/2$ and $2/3$ (since the expansion is assumed to be 
faster than in the radiation dominated case, and is upper limited by the value
for matter dominated expansion).

Note here that using our ansatz for the subdomain scale factors given
by Eq.\eqref{eq:8}, one may try to determine the global scale factor
through Eq.\eqref{globscale}. In order to do so, one needs to know
the inital volume fractions $\lambda_{\ell_{i}}$ which are in turn
related to the $c_{\mathcal{E}}$ and $c_{\mathcal{M}}$. However,
in our approach based upon the Buchert framework \cite{buchert1,buchert2,weigand}
we do not need to determine $c_{\mathcal{E}}$ and $c_{\mathcal{M}}$,
but in stead, obtain from Eq.\eqref{eq:19} the global scale factor
numerically by the method of recursive iteration, using the value
$\lambda_{\mathcal{M}_{0}}=0.09$ determined through numerical simulations
in the earlier literature \cite{weigand}. We later compare these
solutions for the global scale factor with the solutions of the model
with an explicit event horizon studied in Section 4.

\subsection{Backreaction and Scalar Curvature}

It is of interest to study separately the behaviour of the backreaction
term in the Buchert model \cite{buchert1,weigand}. The backreaction
$\mathcal{Q}_{\mathcal{D}}$ is obtained from \eqref{eq:1a} to be
\begin{equation}
\mathcal{Q}_{\mathcal{D}}=3\frac{\ddot{a}_{\mathcal{D}}}{a_{\mathcal{D}}}+4\pi G\left\langle \rho\right\rangle _{\mathcal{D}}\label{eq:14}
\end{equation}
Note that we are not considering the presence of any cosmological
constant $\Lambda$ as shown in \eqref{eq:1a}. We can assume that
$\left\langle \rho\right\rangle _{\mathcal{D}}$ behaves like the
matter energy density, i.e. $\left\langle \rho\right\rangle _{\mathcal{D}}=\frac{c_{\rho}}{a_{\mathcal{D}}^{3}}$,
where $c_{\rho}$ is a constant. Now, observations tell us that the
current matter energy density fraction (baryonic and dark matter)
is about 27\% and that of dark energy is about 73\%. Assuming the
dark energy density to be of the order of $10^{-48}\left(GeV\right)^{4}$,
we get $\rho_{\mathcal{D}_{0}}\backsimeq3.699\times10^{-49}\left(GeV\right)^{4}$.
Thus, using the values for the global acceleration computed numerically,
the future evolution of the backreaction term $\mathcal{Q}_{\mathcal{D}}$
can also be computed (see curves (ii) and (iv) of Fig.2, where we
have plotted the backreaction density fraction $\Omega_{\mathcal{Q}}^{\mathcal{D}}=-\frac{\mathcal{Q}_{\mathcal{D}}}{6H_{\mathcal{D}}^{2}}$).
Once we compute the backreaction it is straightforward to calculate
the scalar curvature $\mathcal{\left\langle R\right\rangle }_{\mathcal{D}}$
as 
\begin{equation}
\mathcal{\left\langle R\right\rangle }_{\mathcal{D}}=16\pi G\left\langle \rho\right\rangle _{\mathcal{D}}-\mathcal{Q}_{\mathcal{D}}-6H_{\mathcal{D}}^{2}\label{eq:15}
\end{equation}

\begin{figure}
\centerline{\includegraphics{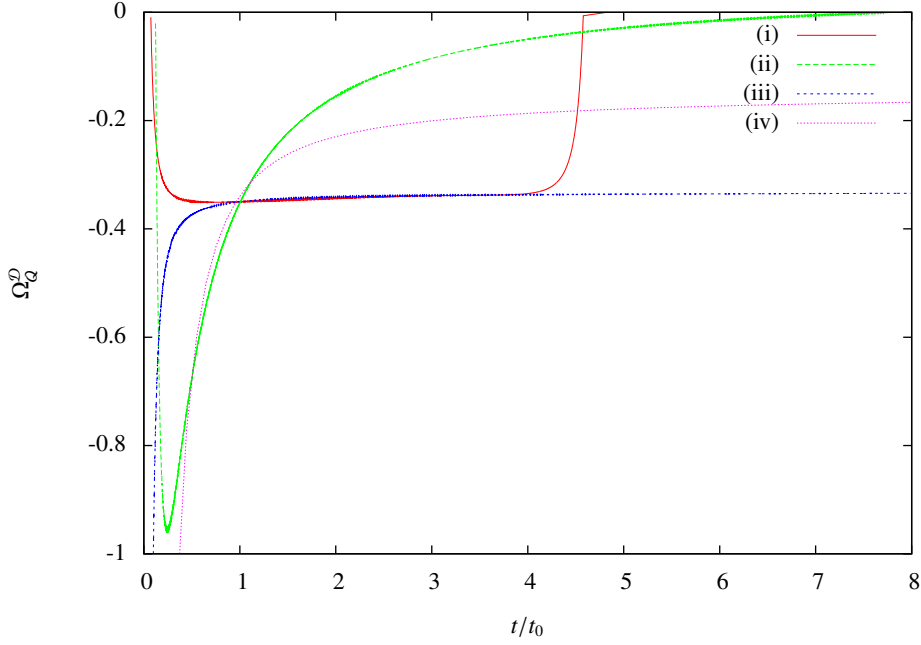} }

\caption{Global backreaction density fraction vs. time. The parameter values
used are the same as in Fig. 1.}

\label{fig2} 
\end{figure}

\begin{figure}
\centerline{\includegraphics{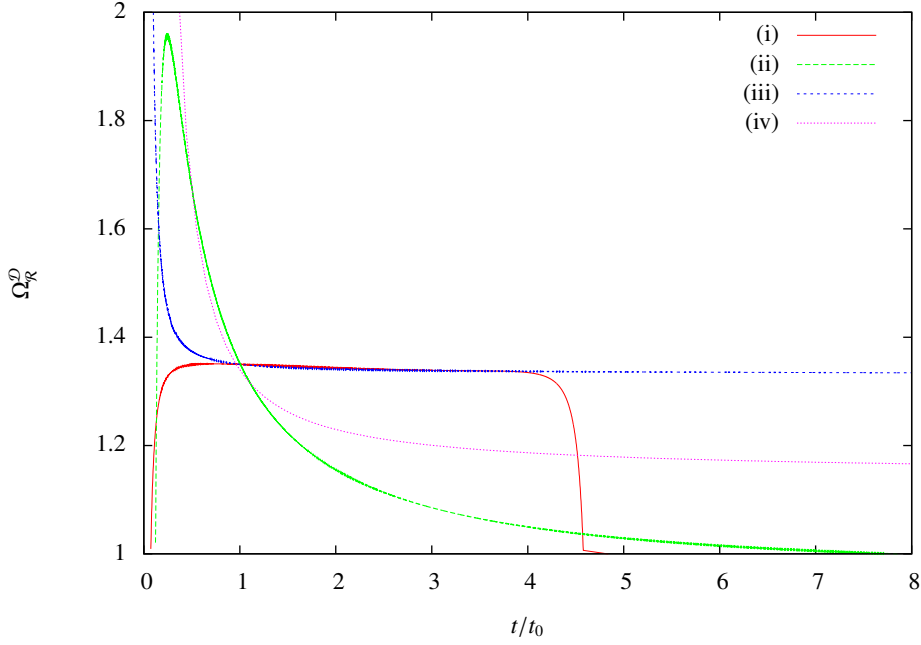} }

\caption{Global scalar curvature density fraction, $\Omega_{\mathcal{R}}^{\mathcal{D}}$
plotted vs. time. The parameter values for curves (i) and (ii) are
$\alpha=0.995$, $\beta=0.5$ , and for curves (iii) and (iv) are
$\alpha=1.02$, $\beta=0.66$. }

\label{fig3} 
\end{figure}

\subsection{Effective Equation of State}

To point out the analogy with the Friedmann equations one may write
\eqref{eq:1a} and \eqref{eq:1b} as 
\begin{eqnarray}
3\frac{\ddot{a}_{\mathcal{D}}}{a_{\mathcal{D}}} & = & -4\pi G\left(\rho_{eff}^{\mathcal{D}}+3p_{eff}^{\mathcal{D}}\right)\nonumber \\
3H_{\mathcal{D}}^{2} & = & 8\pi G\rho_{eff}^{\mathcal{D}}\label{eq:16}
\end{eqnarray}
where the effective energy density and pressure are defined as 
\begin{eqnarray}
\rho_{eff}^{\mathcal{D}} & = & \left\langle \rho\right\rangle _{\mathcal{D}}-\frac{1}{16\pi G}\mathcal{Q}_{\mathcal{D}}-\frac{1}{16\pi G}\mathcal{\left\langle R\right\rangle }_{\mathcal{D}}\nonumber \\
p_{eff}^{\mathcal{D}} & = & -\frac{1}{16\pi G}\mathcal{Q}_{\mathcal{D}}+\frac{1}{48\pi G}\mathcal{\left\langle R\right\rangle }_{\mathcal{D}}\label{eq:17}
\end{eqnarray}
Note that as in \eqref{eq:14}, here also we are not considering the
presence of any cosmological constant $\Lambda$. In this sense, $\mathcal{Q}_{\mathcal{D}}$
and $\mathcal{\left\langle R\right\rangle }_{\mathcal{D}}$ may be
combined to some kind of dark fluid component that is commonly referred
to as $X$-matter. One quantity characterizing this $X$-matter is
its equation of state given by 
\begin{equation}
w_{\Lambda,eff}^{\mathcal{D}}=\frac{p_{eff}^{\mathcal{D}}}{\rho_{eff}^{\mathcal{D}}-\left\langle \rho\right\rangle _{\mathcal{D}}}=\frac{\mathcal{Q}_{\mathcal{D}}-\frac{1}{3}\mathcal{\left\langle R\right\rangle }_{\mathcal{D}}}{\mathcal{Q}_{\mathcal{D}}+\mathcal{\left\langle R\right\rangle }_{\mathcal{D}}}=\frac{\Omega_{\mathcal{Q}}^{\mathcal{D}}-\frac{1}{3}\Omega_{\mathcal{R}}^{\mathcal{D}}}{\Omega_{\mathcal{Q}}^{\mathcal{D}}+\Omega_{\mathcal{R}}^{\mathcal{D}}}\label{eq:18}
\end{equation}
which is an effective one due to the fact that backreaction and curvature
give rise to effective energy density and pressure. We again plot
this effective equation of state by computing its value numerically
(see curves (ii) and (iv) of Fig.4), and in the next Section we compare
the results with those obtained by considering the effect of an event
horizon.

\begin{figure}
\centerline{\includegraphics{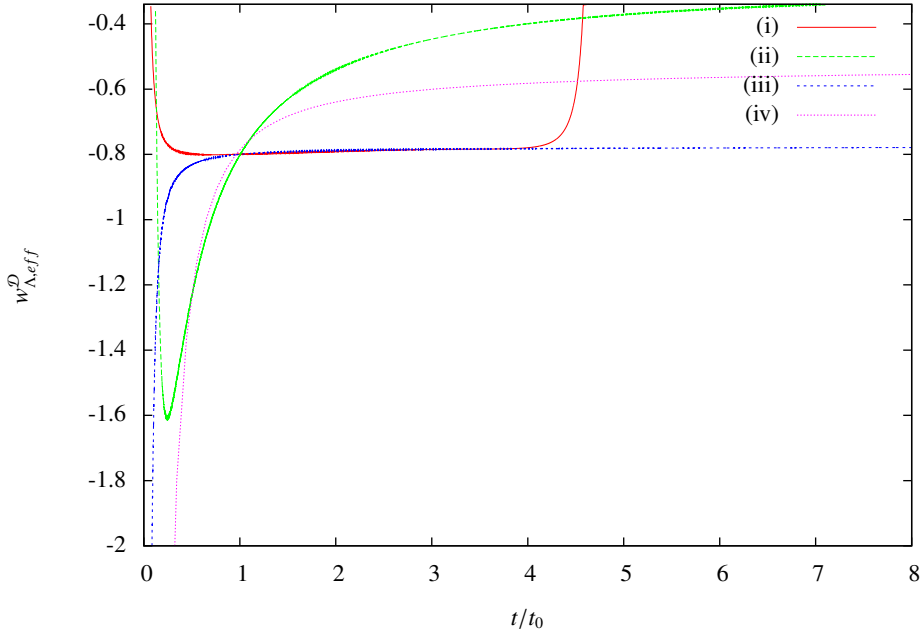} }

\caption{Effective equation of state vs. time. The parameter values used are:
(i) $\alpha=0.995$, $\beta=0.5$, (ii) $\alpha=0.995$, $\beta=0.5,$
(iii) $\alpha=1.02$, $\beta=0.66$, (iv) $\alpha=1.02$, $\beta=0.66$}

\label{fig4} 
\end{figure}

\section{Effect of event horizon}

In the previous section we have performed a detailed analysis of the
future evolution of an accelerating universe by calculating the effect
of backreaction from inhomogeneities on the various dynamical quantities
within the context of the Buchert framework. As mentioned earlier,
an accelerated expansion dictates the formation of an event horizon,
and our earlier analysis may be interpreted as a tacit assumption
of a horizon scale being set by the scale of homogeneity labelled
by the global scale-factor, i.e. $r_{h}\approx a_{\mathcal{D}}$.
In the present section we introduce explicitly such an event horizon.
Given that we are undergoing a stage of acceleration since transition
from an era of structure formation, our aim here is to explore the
subsequent evolution of the Universe due to the effects of backreaction
in presence of the cosmic event horizon. We can write the equation
of the event horizon $r_{h}$, to a good approximation by 
\begin{equation}
r_{h}=a_{\mathcal{D}}\int_{t}^{\infty}\frac{dt'}{a_{\mathcal{D}}(t')}\label{eq:7}
\end{equation}
Note that though, in general, spatial and light cone distances and
corresponding accelerations could be different, as shown explicitly
in the framework of Lemaitre\textendash{}Tolman\textendash{}Bondi
(LTB) models \cite{bolejko}, an approximation for the event horizon
which forms at the onset of acceleration could be defined in the way
above in the same spirit as $a_{\mathcal{D}}=c_{F}a_{F}$. The concept
of the event horizon just ensures that the effect of backreactions
are computed by taking into account only the causally connected processes,
but leaving out the processes that are not causally connected, i.e.,
the effect of inhomogeneities from regions that lie outside the event
horizon. Now, from its very definition the event horizon is observer
dependent. For example, the event horizon for an observer `A' based,
say, in our group of local galaxies, is different from the event horizon
for another hypothetical observer `B' based, say, somewhere in a very
distant region of the universe. This means that certain regions in
the universe that are causally connected to `A' may not be connected
to `B', and vice-versa. The regions not causally connected to `A'
have no impact on the physics, i.e., the spacetime metric for `A'
is unaffected by the backreaction from inhomogeneities at those regions.
Hence, in a two scale void-wall model that we are using, the event
horizon has to be chosen with respect to either `A' (say, wall), or
`B' (say, void). However, the important assumption here is that there
is indeed a scale of global homogeneity which lies within the horizon
volume, and the physics is translationally invariant over such large
scales. The void-wall symmetry of Eq.(\ref{eq:6}) thereby ensures
that the conclusions are similar whether one chooses to define the
event horizon with respect to the wall or with respect to the void.

Since an event horizon forms, only those regions of $\mathcal{D}$
that are within the event horizon are causally accessible to us. 
We hence define a new fiducial global domain as that contained within
the horizon, which naturally is smaller than the original global domain
that we dealt with in our earlier equations. We assume that the entire
Buchert formalism, as outlined in Section 2 holds in this new global 
domain. Note that even if conservation of total rest mass is not 
strictly or exactly obeyed inside this fiducial global domain, the 
magnitude of violation is assumed to be rather small since the volume 
and mass contained within the event horizon is huge, and inflow or 
outflow is assumed to be a rather insignificant fraction of the total 
amount. Therefore, in our subsequent analysis we work under the assumption 
that the Buchert framework is valid up to this approximation.
Denoting this domain as $\widetilde{\mathcal{D}}$ and the
corresponding volume as $\left|\widetilde{\mathcal{D}}\right|_{g}$,
the volume scale-factor is defined as

\begin{equation}
a_{\widetilde{\mathcal{D}}}^{3}=\frac{\left|\widetilde{\mathcal{D}}\right|_{g}}{\left|\widetilde{\mathcal{D}}_{i}\right|_{g}}=\frac{\frac{4}{3}\pi r_{h}^{3}}{\left|\widetilde{\mathcal{D}}_{i}\right|_{g}}\label{eq:9}
\end{equation}
where $\left|\widetilde{\mathcal{D}}_{i}\right|_{g}$ is the volume
of the fiducial global domain at some initial time, which we can take to
be the time when the transition from deceleration to acceleration
occurs. The average of a scalar valued function (eq. \eqref{eq:2})
in $\widetilde{\mathcal{D}}$ will be written as 

\begin{equation}
\left\langle f\right\rangle {}_{\mathcal{\widetilde{D}}}(t)=\frac{\int_{\mathcal{\widetilde{D}}}f(t,X^{1},X^{2},X^{3})d\mu_{g}}{\int_{\mathcal{\widetilde{D}}}d\mu_{g}}=|\mathcal{\widetilde{D}}|_{g}^{-1}\int_{\mathcal{\widetilde{D}}}fd\mu_{g}
\end{equation}

The Einstein equations \eqref{eq:1a}, \eqref{eq:1b}and \eqref{eq:1c}
are also assumed to hold in this new domain, after we replace $\mathcal{D}$
with $\widetilde{\mathcal{D}}$ in the equations. The domain $\widetilde{\mathcal{D}}$
is considered to be divided into several subregions and the average
of the scalar valued function $f$ on the domain $\mathcal{\widetilde{D}}$
may then be split into the averages of $f$ on the subregions $\mathcal{\widetilde{F}}_{\ell}$
in the form, 
\begin{equation}
\left\langle f\right\rangle _{\mathcal{\widetilde{D}}}=\underset{\ell}{\sum}|\mathcal{\widetilde{D}}|_{g}^{-1}\underset{\alpha}{\sum}\int_{\mathcal{\widetilde{F}}_{\ell}^{(\alpha)}}fd\mu_{g}=\underset{\ell}{\sum}\lambda_{\ell}\left\langle f\right\rangle _{\mathcal{\widetilde{F}}_{\ell}}
\end{equation}

where $\lambda_{\ell}=|\mathcal{\widetilde{F}}_{\ell}|_{g}/|\mathcal{\widetilde{D}}|_{g}$,
is the volume fraction of the subregion $\mathcal{\widetilde{F}}_{\ell}$.
Just like in Section 3 here also we consider the global domain $\widetilde{\mathcal{D}}$
to be divided into a collection of overdense regions $\mathcal{M}=\cup_{j}\mathcal{M}^{j}$,
with total volume $|\mathcal{M}|_{g}=\sum_{j}|\mathcal{M}^{j}|_{g}$
, and underdense regions $\mathcal{E}=\cup_{j}\mathcal{E}^{j}$ with
total volume $|\mathcal{E}|_{g}=\sum_{j}|\mathcal{E}^{j}|_{g}$. We also 
assume that the scale-factors of the regions $\mathcal{E}$
and $\mathcal{M}$ are, respectively, given by $a_{\mathcal{E}}=c_{\mathcal{E}}t^{\alpha}$
and $a_{\mathcal{M}}=c_{\mathcal{M}}t^{\beta}$ where $\alpha$, $\beta$,
$c_{\mathcal{E}}$ and $c_{\mathcal{M}}$ are constants. The volume
fraction of the subdomain $\mathcal{M}$ is given by $\lambda_{\mathcal{M}}=\frac{|\mathcal{M}|_{g}}{|\mathcal{\widetilde{D}}|_{g}}$,
which can be rewritten in terms of the corresponding scale factors
as $\lambda_{\mathcal{M}}=\frac{a_{\mathcal{M}}^{3}|\mathcal{M}_{i}|_{g}}{a_{\widetilde{\mathcal{D}}}^{3}|\widetilde{\mathcal{D}}_{i}|_{g}}$. 

We can then find the acceleration of this fiducial global domain $\widetilde{\mathcal{D}}$,
just like we did in Section 3. So in this case the global acceleration
for $\widetilde{\mathcal{D}}$ is given by
\begin{eqnarray}
\frac{\ddot{a}_{\mathcal{\widetilde{\mathcal{D}}}}}{a_{\widetilde{\mathcal{D}}}} & = & \frac{\widetilde{c}_{\mathcal{M}}^{3}t^{3\beta}}{a_{\widetilde{\mathcal{D}}}^{3}}\frac{\beta(\beta-1)}{t^{2}}+\left(1-\frac{\widetilde{c}_{\mathcal{M}}^{3}t^{3\beta}}{a_{\widetilde{\mathcal{D}}}^{3}}\right)\frac{\alpha(\alpha-1)}{t^{2}}\nonumber \\
 &  & +2\frac{\widetilde{c}_{\mathcal{M}}^{3}t^{3\beta}}{a_{\widetilde{\mathcal{D}}}^{3}}\left(1-\frac{\widetilde{c}_{\mathcal{M}}^{3}t^{3\beta}}{a_{\widetilde{\mathcal{D}}}^{3}}\right)\left(\frac{\beta}{t}-\frac{\alpha}{t}\right)^{2}\label{eq:11}
\end{eqnarray}
But we can see from \eqref{eq:9} that $a_{\widetilde{\mathcal{D}}}\propto r_{h}$,
so we can write $\frac{\ddot{a}_{\mathcal{\widetilde{\mathcal{D}}}}}{a_{\widetilde{\mathcal{D}}}}=\frac{\ddot{r}_{h}}{r_{h}}$
and hence the above equation can be written as
\begin{eqnarray}
\frac{\ddot{r}_{h}}{r_{h}} & = & \frac{c_{\mathcal{M}_{h}}^{3}t^{3\beta}}{r_{h}^{3}}\frac{\beta(\beta-1)}{t^{2}}+\left(1-\frac{c_{\mathcal{M}_{h}}^{3}t^{3\beta}}{r_{h}^{3}}\right)\frac{\alpha(\alpha-1)}{t^{2}}\nonumber \\
 &  & +2\frac{c_{\mathcal{M}_{h}}^{3}t^{3\beta}}{r_{h}^{3}}\left(1-\frac{c_{\mathcal{M}_{h}}^{3}t^{3\beta}}{r_{h}^{3}}\right)\left(\frac{\beta}{t}-\frac{\alpha}{t}\right)^{2}\label{eq:11-1}
\end{eqnarray}
Note here that the global domain $\widetilde{\mathcal{D}}$ is a fiducial one
defined with respect to the observer-dependent event horizon for the purpose
of our intermediate analysis, and ultimately we are interested to obtain
the evolution of the global domain $\mathcal{D}$.
In order to get the acceleration of the global domain $\mathcal{D}$
we first convert \eqref{eq:7} to a differential form,
\begin{equation}
\dot{r_{h}}=\frac{\dot{a}_{\mathcal{D}}}{a_{\mathcal{D}}}r_{h}-1\label{eq:13}
\end{equation}
Thus, the evolution of the scale-factor $a_{\mathcal{D}}$ is now
governed by the set of coupled differential equations \eqref{eq:13}
and \eqref{eq:11-1}. We numerically integrate these equations by
using as an `initial condition' the observational constraint $q_{0}=-0.7$,
where $q_{0}$ is the current value of the deceleration parameter. The 
global acceleration obtained from these equations is plotted in Fig. 1 (curves (i) and (iii)).
The expression for $q_{0}$ is a completely analytic function of $\alpha$, $\beta$ 
and $t_{0}$, but we since we are studying the effect of inhomogeneities therefore
the Universe cannot strictly be described based on a FRW model and hence the
current age of the Universe ($t_{0}$) cannot be fixed based on current observations
which use the FRW model to fix the age. Instead for each combination of values of
the parameters $\alpha$ and $\beta$ we find out the value of $t_{0}$ for our model from
\eqref{eq:11} by taking $q_{0}=0.7$. Note that we use the same technique for finding $t_{0}$
in Section 3, where we use \eqref{eq:19}.

\section{Discussions}

Let us now compare the nature of acceleration of the Universe for
the two models described respectively, in Sections 3 and 4. The global
acceleration for the two models have been plotted in Fig. 1. Here
curves (i) and (iii) are for the case when an event horizon is included,
and curves (ii) and (iv) correspond to the case without an event horizon.
For $\alpha<1$ the acceleration becomes negative in the future for
both the cases. The acceleration reaches a much greater value when
no event horizon is present and that too very quickly, but decreases
much more gradually than when an event horizon is included. This behaviour
could be due to the fact that the inclusion of the event horizon somehow
limits the global volume of domain $\mathcal{D}$ in such a way that
the available volume of the underdense region $\mathcal{E}$ is lesser
than when an event horizon is not included. This causes the overdense
region $\mathcal{M}$ to start dominating much earlier and thus causing
global deceleration much more quickly. When $\alpha>1$ the acceleration
decreases and reaches a positive constant value asymptotically when an event
horizon is not included (curve (iv)), and also when it is included
(curve (iii)). Here also we can see that when we include an event
horizon in our calculations then there is much more rapid deceleration,
the reason for which is mentioned above. Initially the curves for
the two cases are almost identical qualitatively, but later on they
diverge due to the faster deceleration for the case when an event
horizon is included.

A similar comparison of the backreaction for the two models is presented
in Fig. 2. From the expression of $\Omega_{\mathcal{Q}}^{\mathcal{D}}$
and \eqref{eq:3} it can be seen that the backreaction will be dominated
by the variance of the local expansion rate $\theta$. Here we observe
that even for the model where an event horizon is not included, $\Omega_{\mathcal{Q}}^{\mathcal{D}}$
is negative for the duration over which the acceleration is positive.
For $\alpha<1$ the backreaction density first reaches a minimum and
then keeps on rising. For the case where an event horizon is included
(curve (i)) the backreaction density remains almost contant after
reaching a minimum but then when it rises it does so very rapidly
as compared to the case where an even thorizon is not included (curve
(ii)). For $\alpha>1$ the backreaction density keeps on rising monotonically.
Here curves for both the models are initially very similar qualitatively
but later on they diverge.

When we look at Fig. 3, where the scalar curvature density has been
plotted we observe that the scalar curvature turns out to be negative
for the duration when the Universe is in accelerated phase, which
is what we expect from our knowledge of FRW cosmology. For $\alpha<1$
the curvature density increases to a maximum and then decreases, the
rise to the maximum being much more fast for the case where an event
horizon is not included (curve (ii)) whereas the fall from the maximum
being much more fast for the case where the event horizon is included
(curve (i)). For $\alpha>1$ the curvature density keeps on decreasing
monotonically for both the cases.

We next consider the effective equation of state $w_{\Lambda,eff}^{\mathcal{D}}$
(Fig. 4). It remains negative for the entire duration over which the
acceleration is positive. For $\alpha<1$, $w_{\Lambda,eff}^{\mathcal{D}}$
first reaches a minimum and then keeps on rising. The falling to the
minimum is much faster for the model where an event horizon is not
included (curve (ii)) whereas for the case with an event horizon the
equation of state remains almost constant after reaching the minimum
but then rises very rapidly (curve(i)). For $\alpha>1$ the curves
for the two cases are very similar initially qualitatively (again
like backreaction plots) but later on they diverge.

\begin{figure}
\centerline{\includegraphics{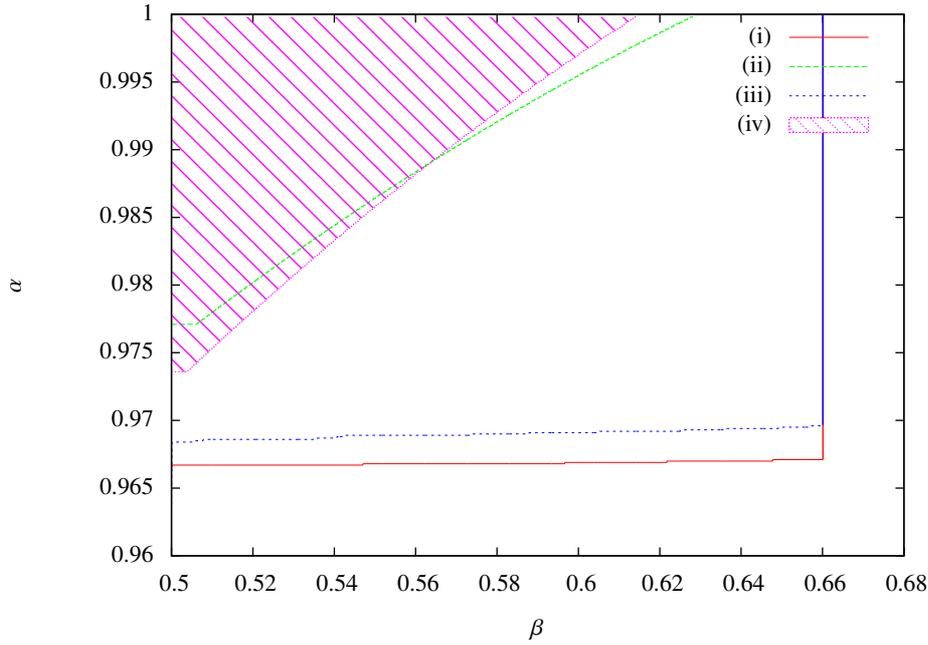} }

\caption{The range of parameters $\alpha$ and $\beta$ for which future deceleration
takes place, shown within the respective contours for the curves (i)
and (iii) corresponding to the case with an event horizon, and the
curves (ii) and (iv) corresponding to the case where an event horizon
is not considered. The value of $\lambda_{\mathcal{M}_{0}}$ for the
curves (i) and (ii) is $0.15$ and for curves (iii) and (iv) is $0.2$.
The shaded region corresponds to the curve (iv) demarcating the parameter
space for this case when acceleration vanishes in finite future time.}

\label{fig5} 
\end{figure}

From our analysis so far it is clear that the acceleration of the
Universe could become negative in the future for certain values of
the parameters $\alpha$ and $\beta$, which represent the growth
rates of the scale factors corresponding to the void and wall, respectively.
The range of values of $\alpha$ and $\beta$ for which a future transition
to deceleration is possible, is depicted in Fig. 5. We provide a contour
plot of $\alpha$ versus $\beta$ demarcating the range in parameter
space (inside of the contours) for which acceleration vanishes in
finite future time. We see that the curves for the model with an event
horizon (curves (i) and (iii)) have the almost the same value of $\alpha$
for various values of $\beta$. This shows that for this case there
is no dependence on the value of $\beta$ to make the acceleration
negative in the future. For the case where there is no event horizon
(curves (ii) and (iv)), initially the values of $\alpha$ are the same for various
values of $\beta$, but later on the values of $\alpha$ begin to
change and increases as $\beta$ increases. Since the acceleration
has no chance of becoming negative when we have $\alpha>1$, therefore
the maximum limit of $\alpha$ for all the cases is depicted by the
line $\alpha=1$.

\section{Conclusions}

To summarize, in this work we have performed a detailed analysis of
the various aspects of the future evolution of the presently accelerating
universe in the presence of matter inhomogeneities. The backreaction
of inhomogeneities on the global evolution is calculated within the
context of the Buchert framework for a two-scale non-interacting void-wall
model \cite{buchert1,buchert2,buchert3,weigand}. We first analyze
the future evolution using the Buchert framework by computing various
dynamical quantities such as the global acceleration, strength of
backreaction, scalar curvature and equation of state. Though in this
case we do not consider explicitly the effect of the event horizon,
it may be argued that a horizon scale is implicitly set by the scale
of global homogeneity labelled by the global scale-factor. We show
that the Buchert framework allows for the possibility of the global
acceleration vanishing at a finite future time, provided that none
of the subdomains accelerate individually (both $\alpha$ and $\beta$
are less than $1$).

We next consider in detail a model with an explicit event horizon,
first presented in \cite{bose}. The observed present acceleration
of the universe dictates the occurrence of a future event horizon
since the onset of the present accelerating era. It may be noted that
though the event horizon is observer dependent, the symmetry of the
equation \eqref{eq:6} ensures that our analysis would lead to similar
conclusions for a `void' centric observer, as it does for a `wall'
centric one. In \cite{bose} we had shown that the presence
of the cosmic event horizon causes the acceleration to slow down significantly
with time. In the present paper in order to understand better the
underlying physics behind the slowing down of the global acceleration,
we have explored the nature of the global backreaction, scalar curvature
and effective equation of state. We have then provided a quantitative
comparison of the evolution of these dynamical quantities of this
model with the case when an event horizon is not included. 

The approach
of incorporating the effects of the event horizon in the present paper 
is different from that used in \cite{bose}. Here we start with the 
assumption that the
portion of the global domain enclosed by the event horizon is a fiducial
domain in which the original Buchert equations are shown to hold. We have then
calculated the acceleration of this fiducial domain using the Buchert
equations. Finally, we have utilized a relation between the event horizon and
the scale factor of the global domain to find out the acceleration of the
original global domain. In contrast, in \cite{bose}  we had explored the 
effect of the event horizon by
assuming that the global domain was cut off by the horizon, thus leading to
apparent volume fractions which we could observe. We then used these apparent
volume fractions in the Buchert acceleration equation to find the global
acceleration, effectively modifying the original Buchert equation.
The main prediction of possible future
deceleration remains the same using both these approaches, though with 
certain quantitative differences in
the rate of deceleration. The present analysis also clarifies that the
prediction of future deceleration may not be solely attributable to the
event horizon, but may also follow purely within the Buchert model
under certain conditions.

Our analysis shows that, in comparison with the model without an event
horizon, during the subsequent future evolution the global acceleration
decreases more quickly at late times when we include an event horizon. 
The reason
for this effect is that in the latter model an effective reduction
of the volume fraction for the void leads to the overdense region
starting to dominate much earlier and hence, causes faster deceleration
of the universe. We also found that the acceleration does not vanish
in finite time, but in stead goes asymptotically to a constant value
for $\alpha>1$ for both the cases. Nonetheless, when $\alpha>1$,
the curves for acceleration, backreaction, scalar curvature, and effective
equation of state for both the cases are very similar qualitatively and only diverge
later on. We finally demarcate the region in the parameter space of
the growth rates of the void and the wall, where it is possible to
obtain a transition to deceleration in the finite future.

Our results indicate the fascinating possibility of backreaction being
responsible for not only the present acceleration as shown in earlier
works \cite{rasanen1,weigand}, but also leading to a transition to
another decelerated era in the future. Another possibility following
from our analysis is of the Universe currently accelerating due to
a different mechanism \cite{sahni}, but with backreaction \cite{buchert1,buchert3,weigand}
later causing acceleration to slow down. Finally, it may be noted
that the formalism for obtaining backreaction \cite{buchert1,buchert2,buchert3,weigand}
used here relies on averaging on constant time hypersurfaces, and incorporating
effects based on causality within this framework may be open to
criticism. A framework more suitable for exploring such effects could be
a procedure based on light-cone averaging that has been proposed recently
\cite{gasperini}. Though the full implications on the cosmological
scenario of the latter framework are yet to be deduced, our analysis
of including the effect of the event horizon introduces certain elements
of lightcone physics in the study of the role of backreaction from
inhomogeneities on the global cosmological evolution. The present 
analysis may be viewed as an attempt to superimpose the effect of
the event horizon on the Buchert framework, and explore if certain interesting
results could be obtained that are at least qualitatively reliable. These
predictions could be open for scrutiny as and when a rigorous formalism is
developed to deal with such effects.

\end{document}